\documentclass[%
superscriptaddress,
preprint,
 amsmath,amssymb,
 aps,
]{revtex4-1}

\usepackage[letterpaper, margin=1in]{geometry}
\usepackage{graphicx}
\usepackage[space]{grffile}
\usepackage{latexsym}
\usepackage{textcomp}
\usepackage{longtable}
\usepackage{tabulary}
\usepackage{booktabs,array,multirow}
\usepackage{amsfonts,amsmath,amssymb}
\usepackage{natbib}
\usepackage{url}
\usepackage{hyperref}
\hypersetup{colorlinks=false,pdfborder={0 0 0}}
\usepackage{etoolbox}

\newif\iflatexml\latexmlfalse

\AtBeginDocument{\DeclareGraphicsExtensions{.pdf,.PDF,.eps,.EPS,.png,.PNG,.tif,.TIF,.jpg,.JPG,.jpeg,.JPEG}}

\usepackage[utf8]{inputenc}
\usepackage[T2A]{fontenc}
\usepackage[ngerman,polish,english]{babel}

\usepackage{amsmath}
\usepackage{mathtools}

\newcommand{\beginsupplement}{%
        \setcounter{table}{0}
        \renewcommand{\thetable}{S\arabic{table}}%
        \setcounter{figure}{0}
        \renewcommand{\thefigure}{S\arabic{figure}}%
     }
     
\iflatexml

\else
\fi

\begin{document}

\title{Dynamical perturbation theory for eigenvalue problems}

\author{Maseim Kenmoe}
\affiliation{Max Planck Institute for Mathematics in the Sciences, Leipzig, Germany}
\affiliation{Mesoscopic and Multilayer Structures Laboratory, Department of Physics, University of Dschang, Cameroon}

\author{Matteo Smerlak}
\affiliation{Max Planck Institute for Mathematics in the Sciences, Leipzig, Germany}

\author{Anton Zadorin}
\affiliation{Max Planck Institute for Mathematics in the Sciences, Leipzig, Germany}

\begin{abstract}
Many problems in physics, chemistry and other fields are perturbative in
nature,~\emph{i.e.} differ only slightly from related problems with
known solutions. Prominent among these is the eigenvalue perturbation
problem, wherein one seeks the eigenvectors and eigenvalues of a matrix
with small off-diagonal elements. Here we introduce a novel iterative
algorithm to compute these eigenpairs based on fixed-point iteration for an algebraic equation in complex projective space.
We show from explicit and numerical examples that our algorithm
outperforms the usual Rayleigh-Schr\"odinger expansion on three counts.
First, since it is not defined as a power series, its domain of
convergence is not~\emph{a priori} confined~to a disk in the complex
plane; we find that it indeed usually extends beyond the standard perturbative radius of
convergence. Second, it converges at a faster rate than the Rayleigh-Schr\"odinger expansion, \emph{i.e.} fewer iterations are required to reach a given
precision. Third, the (time- and space-) algorithmic complexity of each
iteration does not increase with the order of the approximation,
allowing for higher precision computations. Because this complexity is
merely that of matrix multiplication, our dynamical scheme also scales
better with the size of the matrix than general-purpose eigenvalue
routines such as the shifted QR or divide-and-conquer algorithms.
Whether they are dense, sparse, symmetric or~unsymmetric, we confirm
that dynamical diagonalization quickly outpaces LAPACK drivers as the
size of matrices grows; for the computation of just the dominant
eigenvector, our method converges order of magnitudes faster than the
Arnoldi algorithm implemented in ARPACK.~~%
\end{abstract}%

\maketitle

\section{Introduction}

{\label{321240}}
Computing the eigenvalues and eigenvectors of a matrix that is already
close to being diagonal (or diagonalizable in a known basis) is a
central problem in many branches of applied mathematics. The standard
approach, outlined a century ago by Rayleigh~\cite{rayleigh1894} and
Schr\"odinger~\cite{Schr_dinger_1926}, consists in assuming a power series~in a
perturbation parameter~\(\lambda\) and extracting the
coefficients order by order from the eigenvalue equation. Historically,
this technique played a central role in the development of quantum
mechanics, providing analytical insight into~\emph{e.g.} level splitting
in external fields~\cite{Schr_dinger_1926}, radiative
corrections \cite{Bethe_1947} or localization in disordered
systems~\cite{Anderson_1958}. Suitably adapted to the problem at
hand~\cite{M_ller_1934,Epstein_1926,1955}, Rayleigh-Schrödinger (RS) perturbation theory
remains the workhorse of many~\emph{ab initio} calculations in quantum
many-body theory~\cite{ostlund1989}~and quantum field
theory~\cite{itzykson2006}. Outside the physical sciences, RS
perturbation theory underlies the quasispecies theory of molecular
evolution~\cite{Eigen_1988}, among countless other applications.~

{The shortcomings of the RS series are well known. First, its explicit
form requires that all unperturbed eigenvalues be simple, else the
perturbation must first be diagonalized in the degenerate eigenspace
through some other method. Second, the number of terms to evaluate grows
with the order of the approximation; to keep track of them, cumbersome
diagrammatic techniques are often required.~Third, the RS series may not
converge for all (or, in infinite dimensions, any) values
of~\(\lambda\)~\cite{Dyson_1952,Kato_1966,Simon_1991}. Indeed, since the RS series
is a~\emph{power~}series, its domain of convergence is confined to a
disk in the complex~\(\lambda\)}-plane: if a singularity exists
at location~\(\lambda_*\), then the RS series will necessarily
diverge for any value~\(\lambda\) such that~\(\vert\lambda\vert >\vert\lambda_*\vert\).
This means that a non-physical (e.g. imaginary) singularity can
contaminate the RS expansion, turning it into a divergent series in
regions where the spectrum is, in fact, analytic. Of course, divergent
power series can still be useful: the partial sums often approaches the
solution rather closely before blowing up; what is more, a wealth of
resummation techniques, such as Borel resummation or Padé approximants,
can be used to extract convergent approximations from the
coefficients~\cite{kleinert2007}.

In this paper we introduce an explicit perturbation scheme for
finite-dimensional (Hermitian or non-Hermitian) eigenvalue problems
alternative to the RS expansion. Instead of assuming a power series
in~\(\lambda\), we build a sequence of approximations of the
eigenpairs by iterating a non-linear map in matrix space,~\emph{i.e.}
through a discrete-time dynamical system. As we will see, this
alternative formulation of perturbation theory compares favorably with
the conventional RS series---which can be recovered through
truncation---with regards to the second (complexity) and third
(convergence) points above. Our novel scheme is also surprisingly fast
from a computational viewpoint, offering an efficient alternative to
standard diagonalization algorithms for large perturbative matrices.~~

Our presentation combines theoretical results with numerical evidence.
We prove that~\emph{(i)} our dynamical scheme reduces to the~RS partial
sum when truncated to~\(\mathcal{O}(\lambda^k)\), and~\emph{(ii)} there exists a
neighborhood of~\(\lambda=0\) where it converges exponentially to
the full set of perturbed eigenvectors. The latter result is analogous to
Kato's lower bound on the radius of convergence of RS perturbation
theory~\cite{Kato_1966}; like the latter, it provides sufficient but
not necessary conditions for convergence at finite~\(\lambda\).
To illustrate the performance of our algorithm, we analyze several
examples of perturbative matrix families, including a quantum harmonic
oscillator perturbed by a~\(\delta\)-potential and various
ensembles of (dense and sparse) random perturbations.

\begin{figure*}
\begin{center}
\includegraphics[width=1.00\columnwidth]{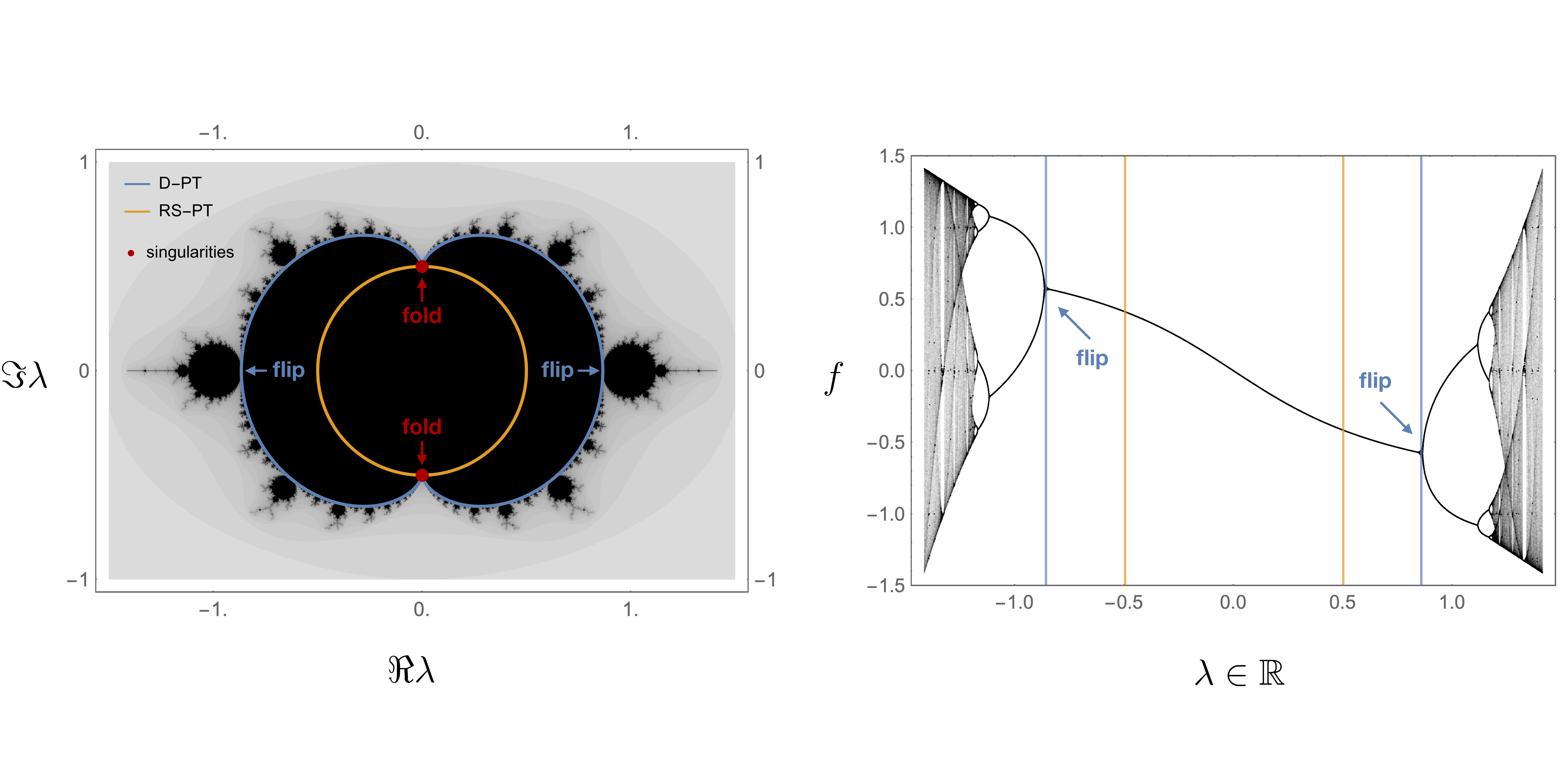}
\caption{{Convergence of perturbation theory in the two-dimensional
example~{\eqref{2by2}}. Left: In the
complex~\(\lambda\)-plane, RS perturbation theory (RS-PT)
converges inside a circle of radius~\(1/2\) (orange line)
bounded by the exceptional points~\(\pm i/2\) where eigenvalues
have branch-point singularities and~\(M\) is not
diagonalizable. Dynamical perturbation theory (D-PT) converges inside
the domain bounded by the blue cardioid, which is larger---especially along the
real axis, where there is no singularity. Outside this domain, the map
can converge to a periodic cycle, be chaotic or diverge to infinity, following flip bifurcations (along the real axis)
and fold bifurcations (at the singularities). The domain where the map
remains bounded (black area) is a conformal transformation of the
Mandelbrot set. Right: The bifurcation diagram for the quadratic
map~\(f\) along the real~\(\lambda\)-axis
illustrates the period-doubling route to chaos as~\(\lambda\)
increases away from~\(0\) (in absolute value). Orange and
left vertical lines indicate the boundary of the convergence domains of
RS-PT and D-PT respectively.~
{\label{907982}}%
}}
\end{center}
\end{figure*}

\section{Results}

{\label{737467}}

\subsection{Eigenvectors as affine
variety}
{\label{185413}}

We begin with the observation that, being defined up to a multiplicative
constant, the eigenvectors of an~\(N\times N\)
matrix~\(M\)~are naturally viewed as elements of the
complex projective space~\(\mathbb{C}P^{N-1}\). Let~\([x^1\,:\,\cdots\, :\, x^N]\)
denote the homogeneous coordinates of~\(x\in\mathbb{C}P^{N-1}\). Fix an
index~\(1\leq n\leq N\), and consider an eigenvector~\(z_n\)
of~\(M\) such that~\(z_n^{\,n}\neq 0\). From the eigenvalue
equation~\(Mz_n=\varepsilon_n z_n\) we can extract the eigenvalue
as~\(\varepsilon_n=(Mz_n)^n/z_n^{\,n}\); inserting this back into~\(Mz_n=\varepsilon_n z_n\)~shows that~\(z_n\)~is a projective root of the system
of~\(N-1\) homogeneous quadratic
equations
$$
(Mz_n)^mz_n^{\,n}=(Mz_n)^nz_n^{\,m}\quad\textrm{for}\quad m\neq n.
$$
Denoting~\(V_n\) the projective variety defined by these
equations and~\(U_n\) the affine chart in which
the~\(n\)-th homogeneous coordinate is nonzero, we arrive
at the conclusion that the set of eigenvectors of~\(M\) is
the affine variety~\(\cup_n(V_n\cap U_n)\), see
SI\ref{SI_EV} for details. Since eigenvectors
generically have non-zero coordinates in all directions, each one
of~\(V_n\cap U_n\) normally contains the complete set of
eigenvectors.~

\subsection{Eigenvectors as fixed
points}

We now further assume a perturbative partioning~\(M=D+\lambda\Delta,\)
where the diagonal part~\(D\) consists of unperturbed
eigenvalues~\(\epsilon_n\) and~\(\lambda\Delta\) is a
perturbation parametrized by~\(\lambda\in \mathbb{C}\). Provided
the~\(\epsilon_n\)'s are all simple (non-degenerate), we can
rewrite the polynomial system above
as\[ z_n^{\,n}z_n^{\,m}=\lambda\theta_{n}^{\,m}\left(z_n^{\,n}(\Delta z_n)^m-z_n^{\,m}(\Delta z_n)^n\right) \quad\textrm{for}\quad m\neq n,
 \]where~\(\theta_{n}^{\,m}\equiv (\epsilon_n-\epsilon_m)^{-1}\). Within the
chart~\(U_n\) we are free to set~\(z_n^{\,n}=1\); this
results in the fixed point
equation~\(z_n=F_n(z_n)\)~with~\(F_n:U_n\to U_n\)~the map with
components
\begin{equation}\label{single_line_dyn}
F_n^{\,m}(z_n)\equiv\delta_{n}^{\,m}+\lambda\theta_{n}^{\,m}\left((\Delta z_n)^m-(\Delta z_n)^nz_n^{\,m}\right) \quad \textrm{for}\quad 1\leq m\leq N,
\end{equation}
where the matrix~\(\theta\) is
completed with~\(\theta_{n}^{\, n} = 0\) and \(\delta_{n}^{\,m}\) are the
components of the unit matrix \(I\).

As noted above, each one of these maps~\(F_n:U_n\to U_n\) generically
has the full set of eigenvectors of~\(M\) as fixed points.
Here, however, we wish to compute these fixed points dynamically, as
limits of the iterated map~\[z_n^{(k+1)}=F_n(z_n^{(k)}).
\]For this procedure to
converge, the map~\(F_n\) must be contracting in the
neighborhood of at least one fixed point; for small~\(\lambda\)
this is always the case for the fixed point~closest to the unperturbed
eigenvector~\(e_n\equiv(\delta_{n}^{\,m})_{1\leq m\leq N}\), but usually not for the other fixed
points. To compute all the eigenvectors of~\(M\), we must
therefore iterate not one, but all the
maps~\(F_n\)~together, ideally in parallel.~

We can achieve this by bundling all~\(N\)candidate
eigenvectors in a list and applying the map \(F_n\) to
the~\(n\)-th component:

\begin{align*}
F:\prod_{1\leq n\leq N}U_n&\longrightarrow\prod_{1\leq n\leq N}U_n\\ (z_1,\cdots,z_N)&\longmapsto (F_1(z_1),\cdots,F_N(z_N))
\end{align*}
Introducing the matrix~\(A\) whose~\(n\)-th
line is the vector~\(z_n\), this can be written in matrix
form as
\begin{equation}\label{dyn}
	F(A)=I+\lambda\theta\star\left(A\Delta'-(A\Delta')\triangleright A\right).
\end{equation}
Here prime denotes the
transpose,~\(\star\)~the Hadamard (element-wise) product of
matrices, and~\(A\triangleright B\equiv (I\star A)B\). The dynamical system in the title is
then obtained through iterated applications of~\(F\) to the
identity matrix of unperturbed eigenvectors~\(I\),
generating the sequence of approximating eigenvectors~
\begin{equation}
	A_{\textrm{D}}^{(k)}=F(A_{\textrm{D}}^{(k-1)})=\cdots=F^{\circ k}(I).
\end{equation}
The eigenvalues~\(\varepsilon^{(k)}_n\) at order~\(k\) are
given by~\(\varepsilon_n^{(k)}=\epsilon_n+(A_{\textrm{D}}^{(k)}\Delta')_{n}^{\, n}\).

\subsection{Convergence at small~\(\lambda\)}

{\label{996058}}

It is not hard show that~\(A_{\textrm{D}}^{(k)}\)~always converges at
small~\(\lambda\). Let~\(\Vert\cdot\Vert\) denote the spectral
norm of matrices (the largest singular value), which is
sub-multiplicative with respect to both the matrix and Hadamard
products~\cite{r2012}, and assume~\(\Vert\Delta\Vert=1\) without loss
of generality. First,~\(\Vert F(A)-I\Vert\leq \Vert \lambda \theta\Vert (\Vert A\Vert+\Vert A\Vert^2)\) implies
that~\(F\) maps a closed ball~\(B_r(I)\) of
radius~\(r\) centered on~\(I\) onto itself
whenever~\(\Vert\lambda\theta\Vert\leq r/[(1+r)(2+r)]\).~ Next, from~{\eqref{dyn}}
we have\[\Vert F(A)-F(B)\Vert\leq \Vert\lambda\theta\Vert\,(1+\Vert A+  B\Vert)\, \Vert A -  B\Vert.\]Hence,~\(F\)~is contracting
in~\(B_r(I)\)~provided~\(\Vert\lambda\theta\Vert < 1/[1+2(1+r)]\). Under these
conditions the Banach fixed-point theorem implies
that~\(A_{\textrm{D}}^{(k)}=F^{\circ k}(I)\) converges exponentially to a unique fixed point
within this ball as~\(k\to\infty\). Choosing the optimal
radius\[\underset{r>0}{\textrm{argmax}}\min\,\left(\frac{r}{(1+r)(2+r)},\frac{1}{1+2(1+r)}\right)=\sqrt{2},\]

we see that~\(\vert\lambda\vert<(3-2\sqrt{2})\Vert\theta\Vert^{-1}\approx 0.17\Vert\theta\Vert^{-1}\) guarantees convergence to the fixed
point.~

\subsection{Contrast with RS perturbation theory}

{\label{996058}}

Let us now contrast this iterative method with conventional RS
perturbation theory. There, the eigenvectors of~\(M\) are
sought as power series
in~\(\lambda\),~\emph{viz.}~\(z=\sum_{\ell}z^{(\ell)}\lambda^\ell\). Identifying the
terms of the same order in~\(\lambda\) in the eigenvalue
equation, one obtains for the matrix of eigenvectors at
order~\(k\)\[
A^{(k)}_{\textrm{RS}}=\sum_{\ell=0}^k a^{(\ell)}\lambda^\ell\] in which the matrix
coefficients~\(a^{(\ell)}\)~are obtained from~\(a^{(0)}=I\)
via the recursion
(SI{\ref{SI_RSPT}})
\begin{equation}\label{ARS-alpha}
	a^{(\ell)}=\theta\star\left(a^{(\ell-1)}\Delta'-\sum_{s=0}^{\ell-1}(a^{(s)}\Delta')\triangleright a^{(\ell-1-s)}\right).
\end{equation}
We show in SI{\ref{SI_D_RS_PT}} that the dynamical
scheme~\(A^{(k)}_{\textrm{D}}\)~completely contains this RS series, in the
sense that~\(A^{(k)}_{\textrm{D}}=A^{(k)}_{\textrm{RS}}+\mathcal{O}(\lambda^{k+1}).\) For example, to order~\(2\)
we have~\[A^{(2)}_{\textrm{RS}}=I+\lambda\theta\star\Delta'+\lambda^2\theta\star\Big((\theta\star\Delta')\Delta'\Big),
\]and~\[A^{(2)}_{\textrm{D}}=A^{(2)}_{\textrm{RS}}+\lambda^3\theta\star\Big[\Big((\theta\star\Delta')\Delta'\Big)\triangleright(\theta\star\Delta') \Big].
\]
This means that we can recover the usual perturbative expansion
of~\(A\)~to order~\(k\) by
iterating~\(k\) times the map~\(F\) and
dropping all terms~\(\mathcal{O}(\lambda^{k+1})\). Moreover, the parameter whose
smallness determines the convergence of the RS series is the product of
the perturbation magnitude~\(\lambda\) with the inverse diagonal
gaps~\(\theta\)~\cite{Kato_1966}, just as it determines the
contraction property of~\(F\).~

These similarities notwithstanding, the dynamical scheme~differs from
the RS series in two key ways. First, the complexity of each iteration
is constant (one matrix product with~\(\Delta'\) and one
Hadamard product with~\(\theta\), which is faster), whereas
computing the RS coefficients~\(a^{(\ell)}\)~involves the sum of
increasingly many matrix products. Second, not being defined as a power
series, the convergence of~\(A^{(k)}_{\textrm{D}}\) when~\(k\to\infty\)
is not~\emph{a priori~}restricted to a disk in the
complex~\(\lambda\)-plane. Together, these two differences
suggest that our dynamical scheme has the potential to converge faster,
and in a larger domain, than RS perturbation theory. This is what we now
examine, starting from an elementary but explicit example.

\subsection{An explicit \(2\times 2\)
example}

Consider the symmetric matrix 
\begin{equation}\label{2by2}
M =  \begin{pmatrix}
   0 & 0 \\
   0 & 1 
\end{pmatrix}+\lambda \begin{pmatrix}
   0 & 1 \\
   1 & 0 
\end{pmatrix}.
\end{equation}
This matrix has eigenvalues $\varepsilon_{\pm}=(1\pm\sqrt{1+4\lambda^2})/2$, both of which are analytic inside the disk $\vert\lambda\vert<1/2$ but have branch-point singularities at $\lambda=\pm i/2$. (These singularities are in fact exceptional points, i.e. $M$ is not diagonalizable for these values.) Because the RS series is a power series, these imaginary points contaminate its convergence also on the real axis, where no singularity exists: $A_{\textrm{RS}}$ diverges for any value of $\lambda$ outside the disk of radius $1/2$, and in particular for real $\lambda>1/2$ .

Considering instead our iterative scheme, one easily computes $$A^{(k)}_{\textrm{D}}=\begin{pmatrix}
    1 & f^{\circ k}(0) \\
    -f^{\circ k}(0) & 1
    \end{pmatrix},$$
where $f(x)=\lambda(x^2-1)$ and the superscripts indicate $k$-fold iterates. This one-dimensional map has two fixed points at $x^*_{\pm}=\varepsilon_{\pm}/\lambda$. Of these two fixed points $x^*_+$ is always unstable, while $x^*_-$ is stable for $\lambda \in(-\sqrt{3}/2, \sqrt{3}/2)$ and loses its stability at $\lambda=\pm\sqrt{3}/2$ in a flip bifurcation. At yet larger values of $\lambda$, the iterated map $f^{\circ k}$---hence the fixed-point iterations $A^{(k)}_{\textrm{D}}$---follows the period-doubling route to chaos familiar from the logistic map \cite{May_1976}. For values of $\lambda$ along the imaginary axis, we find that the map is stable if $\Im \lambda\in(-1/2,1/2)$ and loses stability in a fold bifurcation at the exceptional points $\lambda=\pm i/2$. The full domain of convergence of the system is strictly larger than the RS disk, as shown in Fig. \ref{907982}. Further details on the algebraic curve bounding this domain are given in SI\ref{SI_domain}-\ref{SI_examples}. 

We also observe that the disk where both schemes converge, the dynamical scheme does so with a better rate than RS perturbation theory: we check that $\vert f^{\circ k}(0)-x^*_-\vert\sim\vert1-\sqrt{1+4\lambda^2}\vert^k=\mathcal{O}(\vert 2\lambda^2\vert^k)$, while the remainder of the RS decays as $\mathcal{O}(\vert 2\lambda\vert^k)$. This is a third way in which the dynamical scheme outperforms the usual RS series, at least in this case: not only is each iteration computationally cheaper, but the number of iterations required to reach a given precision is lower. 

\subsection{Quantum harmonic oscillator
with~\(\delta\)-potential}

{\label{369545}}

As a second illustration let us consider the Hamiltonian of a quantum
harmonic oscillator perturbed by a~\(\delta\)-potential at the
origin,~
\begin{equation}\label{delta_potential}
	H=\left(-\frac{1}{2}\frac{\partial^2}{\partial x^2}+\frac{1}{2}x^2\right)+\lambda \delta(x),
\end{equation}
where the unperturbed Hamiltonian (in
brackets) is diagonal in the basis of Hermite
functions~\(\phi_{n}(x)=(2^{n} n!\sqrt{\pi})^{-1/2}H_{n}(x)e^{-x^2/2}\) with eigenvalues~\(\epsilon_n=n+1/2\). For
this example familiar from elementary quantum
mechanics~\cite{Viana_Gomes_2011}, it was estimated in~\cite{Kvaal_2011}
that the RS series converges for all~\(\vert\lambda\vert<2\). Since (by
parity) the perturbation only affects the even-number eigenfunctions, we
consider the matrices~\(D_{n}^{\, m}=(2n+1/2)\delta_{n}^{\, m}\) and~\(\Delta_{n}^{\, m}=\phi_{2n}(0)\phi_{2m}(0)\)
with~\(n,m\) truncated at some value~\(N\).~

\begin{figure*}
\begin{center}
\includegraphics[width=.7\columnwidth]{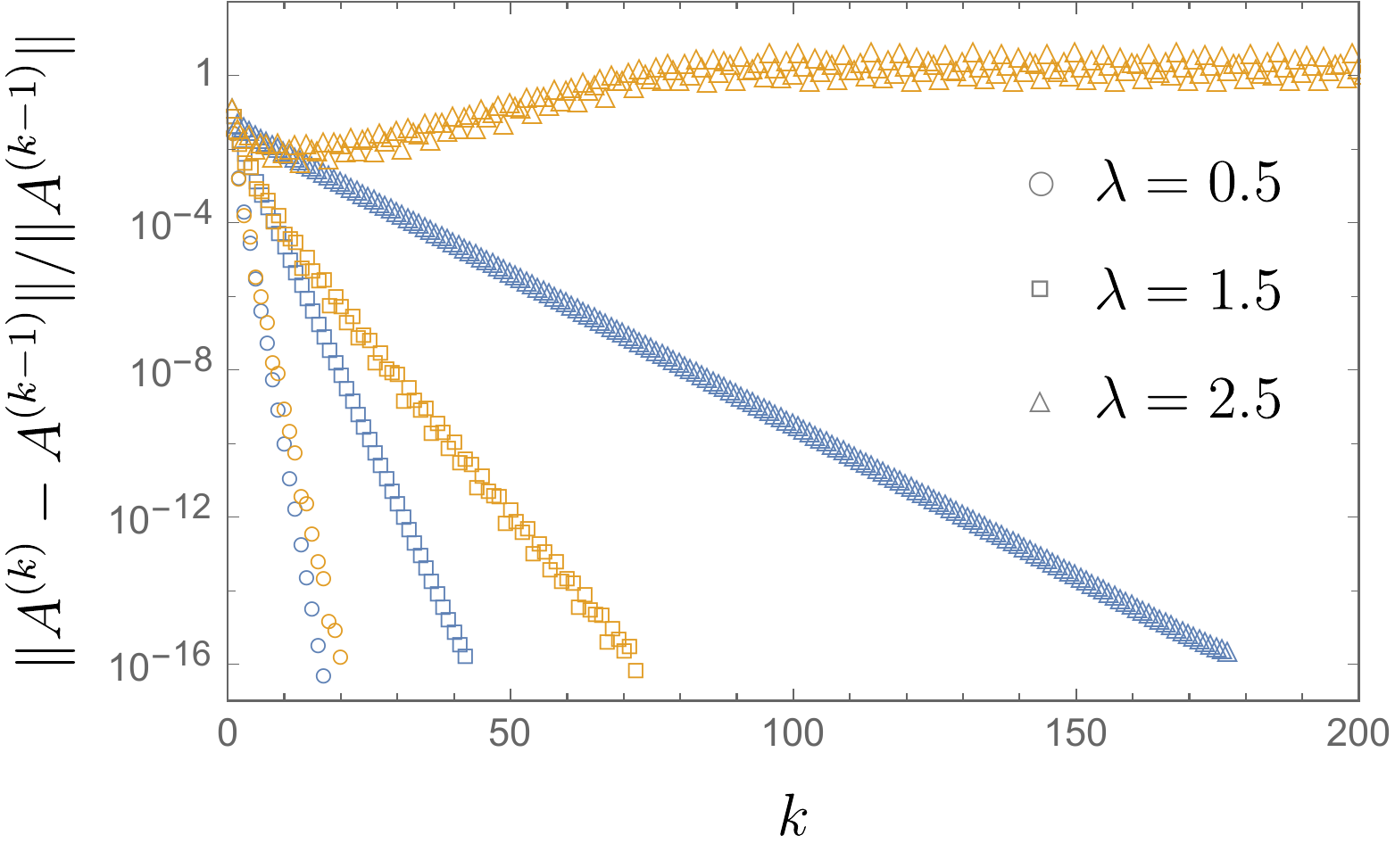}
\caption{{Convergence of the RS perturbative series (RS-PT, orange symbols) and of
our dynamical scheme (D-PT, blue symbols) for the perturbed harmonic
oscillator in~{\eqref{delta_potential}}
with~\(N=100\) and increasing values of~\(\lambda\). The
latter converges at a faster rate, and in a larger domain, than the RS
series.~
{\label{843051}}%
}}
\end{center}
\end{figure*}

Fig.~{\ref{843051}} compares the convergence of the
matrix of eigenvectors~\(A^{(k)}_\textrm{RS,D}\) with the
order~\(k\) under the RS and dynamical schemes. While
convergence is manifestly~ exponential in both cases, the rate is faster
in dynamical perturbation theory, particularly at larger values
of~\(\lambda\). For instance, for~\(\lambda=1.5\) it takes
over~\(70\) orders in conventional perturbation theory to
reach the same precision as~\(\sim40\) iterations of the
map~\(F\); when~\(\lambda=2.5\) the RS series no
longer converges, but the orbits of \eqref{dyn} still
do. In the limit where~\(N\to\infty\) (the complete quantum
mechanical problem) we found that the RS series has a radius of
convergence~\(\approx 2.2\), whereas dynamical perturbation theory
converges up to~\(\lambda_{\textrm{max}}\approx 2.9\).

\subsection{An improved perturbation
theory}

{\label{394233}}

These improved convergence properties are not special to the matrix
families~{\eqref{2by2}}
or~{\eqref{delta_potential}}. To see this, let us now
consider an ensemble of (nonsymmetric) random matrices of the
form~\(M=\textrm{diag}(n)_{1\leq n\leq N}+\lambda R\) where~\(R\)~are~\(N\times N\)
arrays of random numbers uniformly distributed
between~\(-1\) and~\(1\). For
given~\(N\) and~\(\lambda\), we compute the success
rate (probability of convergence) of each method, and, in cases where
both converge, the orders \(K_\textrm{D}\) and~\(K_\textrm{RS}\)
required by each to converge at hundred-digit precision.~

\begin{figure*}
\begin{center}
\includegraphics[width=.8\columnwidth]{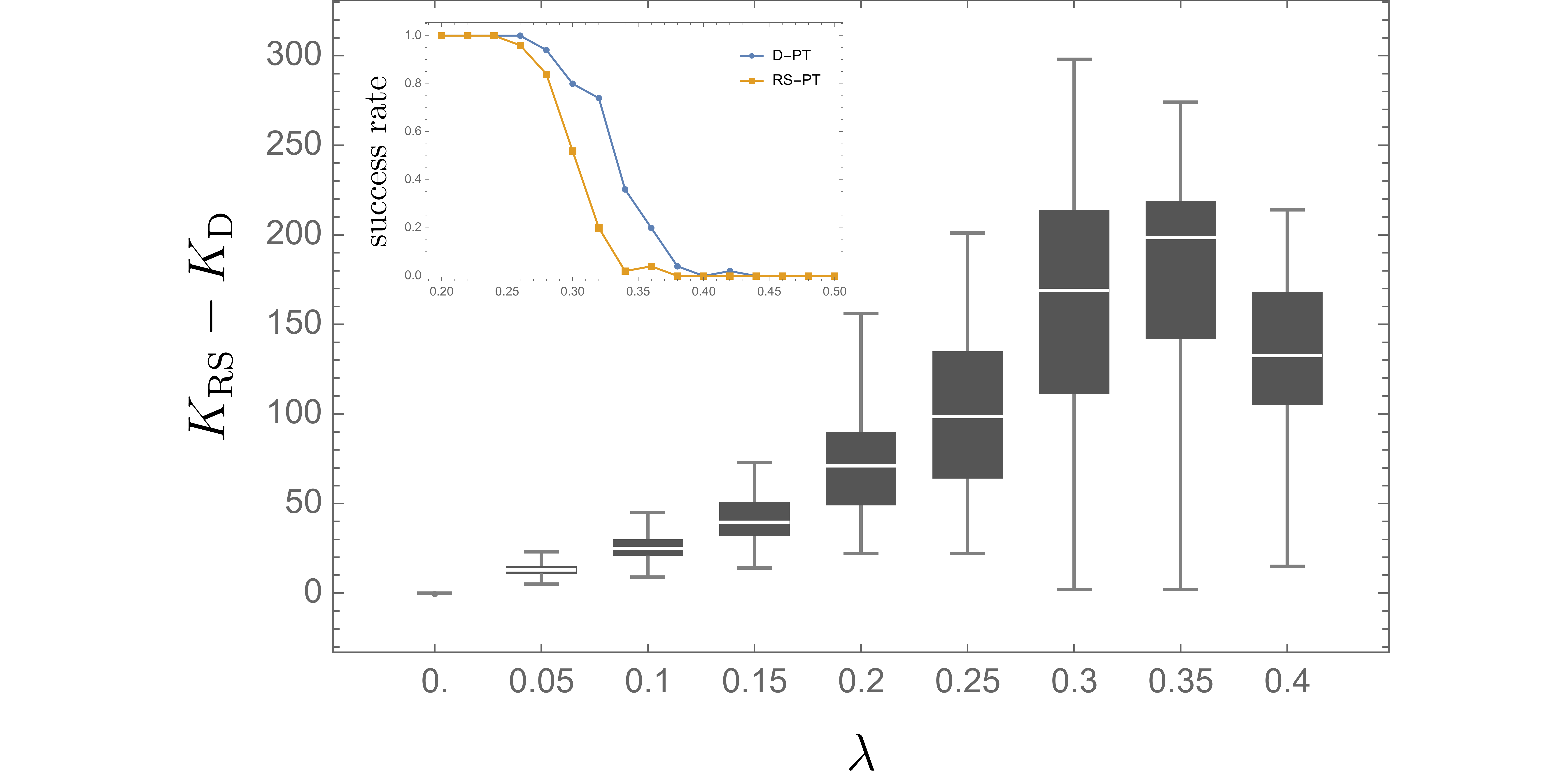}
\caption{{Comparison of the convergence rates of dynamical perturbation theory
(D-PT) and RS perturbation theory (RS-PT) for random matrices
with~\(N=100\), equally spaced diagonal entries and uniformly
distributed off-diagonal elements. A positive value
of~\(K_{\textrm{RS}}-K_{\textrm{D}}\) indicates that D-PT reaches hundred-digit
precision for the full set of eigenvectors with fewer iterations than
RS-PT; this is true in a large majority of cases, though not always. The
inset shows that D-PT is also more likely to converge (for a
given~\(\lambda\)) than RS-PT.~
{\label{161107}}%
}}
\end{center}
\end{figure*}

Fig.~{\ref{161107}} plots the
difference~\(K_\textrm{RS}-K_\textrm{D}\) for matrices of size~\(N=100\)
at various values of~\(\lambda\). In a large majority of cases
(out of~\(50\) samples for each~\(\lambda\)), the
difference turns out positive, implying that dynamical perturbation
converges as a faster rate than RS perturbation theory. The success rate
(inset) is also better with the former, confirming the patterns observed
in the~\(2\times 2\) and~\(\delta\)-potential examples.~

\subsection{A subcubic eigenvalue
algorithm}

{\label{234116}}

We noted earlier that the algorithmic complexity of each iteration of
the map is limited by that of matrix multiplication
by~\(\Delta'\). The complexity of matrix multiplication is
theoretically~\(\mathcal{O}(N^{2.37\cdots})\)~\cite{Coppersmith_1990,Le_Gall_2014}; the Strassen
algorithm~\cite{Strassen_1969}~often used in practice performs
in~\(\mathcal{O}(N^{2.8\cdots})\) or better for sparse
matrices. This implies that, within its domain of convergence (i.e. for
matrices with small off-diagonal elements), our dynamical scheme scales
better than classical diagonalization algorithms such as shifted
Hessenberg QR (for nonsymmetric matrices) or divide-and-conquer (for
Hermitian matrices), which are all~\(\mathcal{O}(N^3)\) if all eigenpairs
are requested~\cite{Demmel_1997}.~

We tested this premise by comparing the timing of the complete dynamical
diagonalization of large matrices with general-purpose LAPACK
routines~\cite{d1999}, implemented in Mathematica 12 in
the function~\textbf{Eigensystem}. To this aim we considered two classes
of random perturbations~of the harmonic oscillator
in~{\eqref{delta_potential}}: dense nonsymmetric
matrices with uniformly distributed entries, and sparse Laplacian (hence
symmetric) matrices of critical Erd\selectlanguage{polish}ő\selectlanguage{english}s-R\'enyi random graphs (i.e.
with~\(N\) vertices and~\(N\) edges). The
perturbation parameter was set to~\(\lambda=0.01\) and timings were
made both at machine precision and at fixed precision
(\(p=100\) digits). Finally, we used a partioning of the
matrices in which~\(\Delta\) has vanishing diagonal elements;
this partioning is known in quantum chemistry as Epstein-Nesbet
perturbation theory~\cite{Epstein_1926,1955}. In each case dynamical
perturbation theory proved (much) faster than LAPACK routines at
large~\(N\).
\begin{figure*}
\begin{center}
\includegraphics[width=1.00\columnwidth]{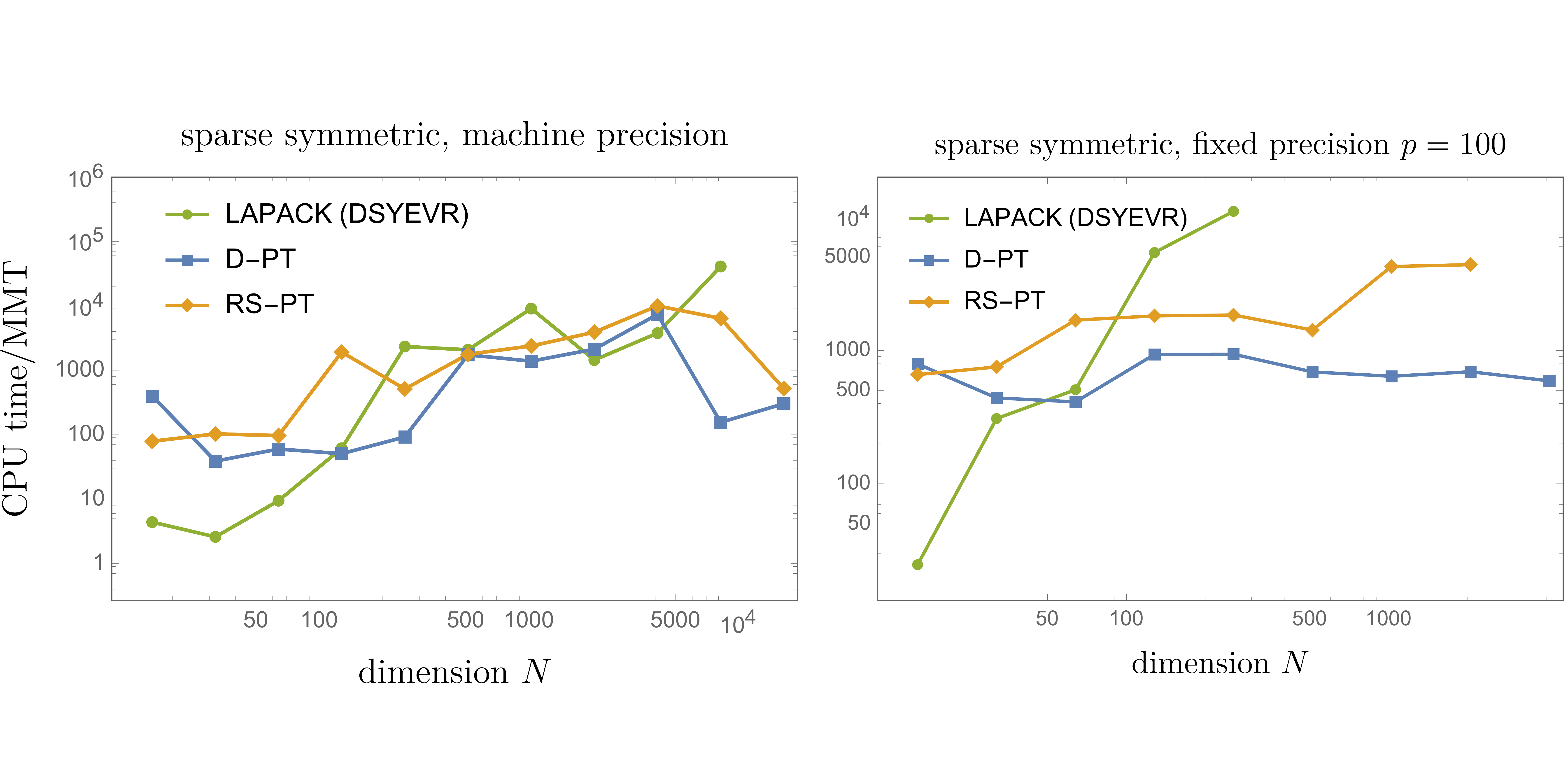}
\caption{Performance of perturbative (dynamical, D-PT, and Rayleigh-Schr\"odinger,
RS-PT) diagonalization vs. the~\textbf{DSYEVR~}LAPACK routine
(implemented as~\textbf{Eigensystem~}in Mathematica 12) for sparse
symmetric matrices, given as ratios of CPU time to matrix multiplication
time (MMT). Unlike standard cubic algorithms, the complexity of our
dynamical scheme is just that matrix multiplication.~ As a result, D-PT
quickly outpaces LAPACK at both machine (left) or---even more
impressively---fixed precision (right). The values given are averages
over~\(50\) repetitions with a time constraint of one hour.~
{\label{861251}}%
}
\end{center}
\end{figure*}

Representative results are given in
Fig.~{\ref{861251}}, where we plot the ratio of the CPU
time required to compute all the eigenvectors of
sparse~\(M\)~to the CPU time to compute its
square~\(M^2\) (matrix multiplication time MMT). While this
ratio increases sharply with~\(N\) with standard
algorithms, it does not with our iterative algorithm. As a result,
dynamical perturbation theory quickly outperforms LAPACK (here
the~\textbf{DSYEVR~}routine for symmetric matrices), and especially its
fixed precision version in Mathematica. The large improvement is
remarkable because symmetric eigenvalue solvers are among the best
understood, and most fine-tuned, algorithms in numerical mathematics \cite{Golub_2000}.
 Our
implementation of~{\eqref{dyn}}, on the other hand, was
written entirely in Mathematica,~\emph{i.e.} without the
benefit of compiled languages.~

\begin{figure*}
\begin{center}
\includegraphics[width=0.70\columnwidth]{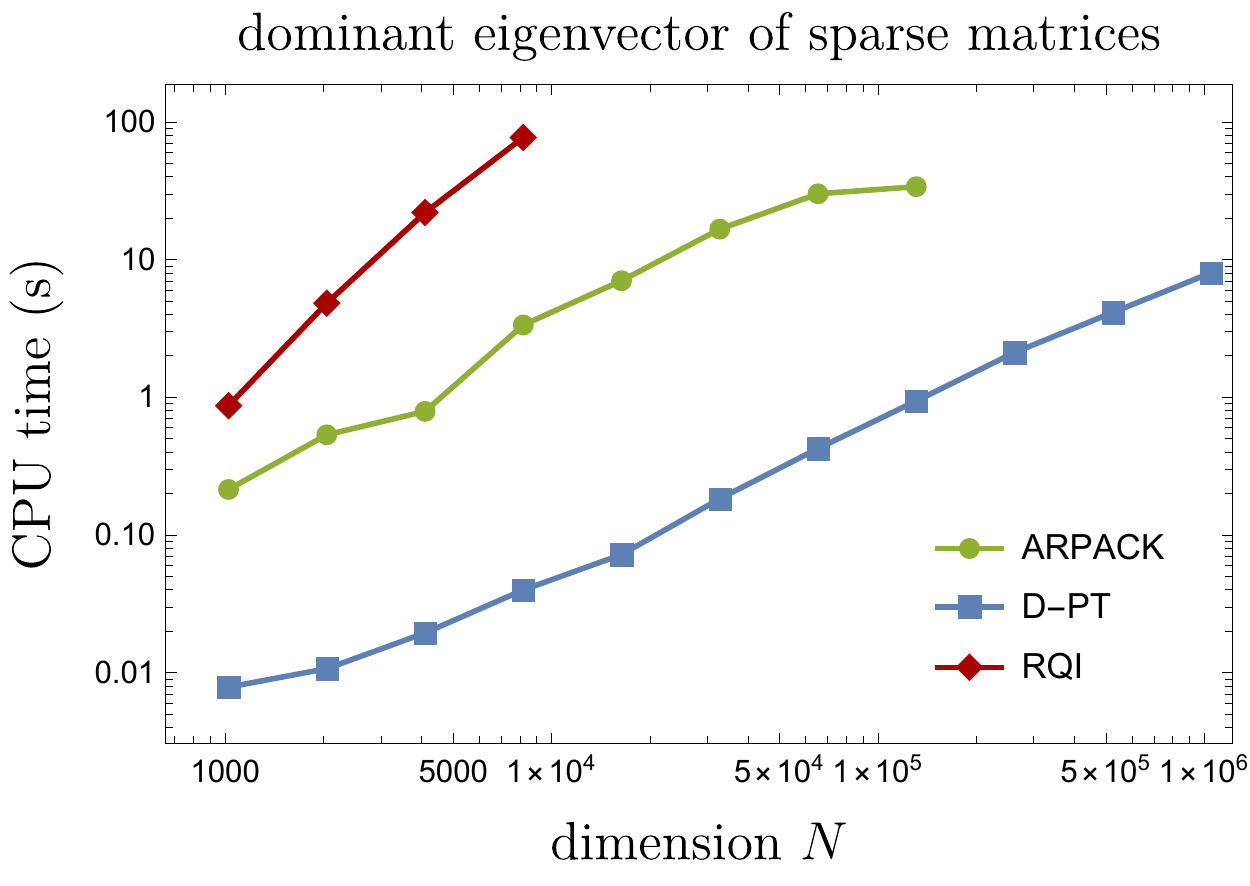}
\caption{CPU time to compute the dominant eigenvector of sparse perturbative
matrices at machine precision with the Lanczos algorithm in
Mathematica's ARPACK routine (with starting vector provided), with the
single-vector map~{\eqref{single_line_dyn}} (D-PT), and with
Reyleigh quotient iteration (RQI). Dynamical perturbation theory is much
faster, even in our non-compiled Mathematica implementation.~
{\label{816420}}%
}
\end{center}
\end{figure*}

Finally we considered the computation of just the dominant eigenvector
of~\(M\) (the one with the largest eigenvalue). In contrast
with the complete eigenproblem, subcubic algorithms exist for finding
such extremal eigenvectors, such as the Arnoldi and Lanczos algorithms
or Rayleigh quotient iteration~\cite{Demmel_1997}. Thanks to their
iterative nature---similar to our dynamical scheme---these algorithms
perform especially well with sparse matrices, and are generally viewed
as setting a high bar for computational efficiency. Yet, for sparse
perturbative matrices (such that the dominant
eigenvector~\(z_{n^*}\)~is the one associated with the largest
unperturbed eigenvalue~\(\epsilon_{n^*}\)), we found that the iteration
of the single-vector map~\(F_{n^*}\) converges orders of
magnitudes faster than these iterative algorithms
(Fig.~{\ref{816420}}). For instance, it took
just~\(4\) iterations and~\(3\) seconds on a
desktop machine to compute at machine precision the dominant eigenpair
of a sparse symmetric matrix with~\(10^8\) non-zero elements
(\(\sim 1.5\) Go in memory); by contrast ARPACK's Lanczos algorithm had not converged within a~\(100\)s
cutoff for a matrix~\(50\) times smaller.~

Overall, these timings suggest that, although certainly not a
general-purpose method due to its limited convergence domain, dynamical
diagonalization can be an efficient approach to the numerical eigenvalue
problem, whether all eigenpairs are needed (in which case it is
pleasantly parallel) or just a few. When the matrix of interest has
off-diagonal elements which are not small, it might still be possible to
compute approximate eigenvectors through another method (say
to~\(1\%\) accuracy) and improve the performance by using
dynamical diagonalization for the final convergent steps.~

\section{Discussion}

{\label{898836}}

We reconsidered the classic problem of computing the eigensystem of a
matrix given in diagonal form plus a small perturbation. Departing from
the usual prescription based on power series expansions, we identified a
quadratic map in matrix space whose fixed points are the sought-for
eigenvectors, and proposed to compute them through fixed-point
iteration,~\emph{i.e.} as a discrete-time~ dynamical system. We showed
empirically that this algorithm outperforms the usual RS series on three
counts: each iteration has lower computational complexity, the rate of
convergence is usually higher (so that fewer iterations are needed to
reach a given precision), and the domain of convergence of the scheme
(with respect to the perturbation parameter~\(\lambda\)) is
usually larger. Within its domain of convergence, our algorithm runs
much faster than standard exact diagonalization routines because its
complexity is merely that of matrix multiplication (by the perturbation
\(\Delta\)). ~

Several extensions of our method can be considered. For instance,
when~\(\lambda\) is too large and the dynamical system does not
converge, we may reduce it to~\(\lambda/Q\) for some
integer~\(Q\), diagonalize the matrix with this smaller
perturbation, restart the algorithm using the diagonal matrix thus
obtained as initial condition, and repeat the
operation~\(Q\) times. This approach is similar to the
homotopy continuation method for finding polynomial
roots~\cite{Morgan_2009} and can effectively extend the domain of
convergence of the dynamical scheme. Another avenue is to leverage the
geometric structure outlined in~ Sec. {\ref{185413}}, for
instance by using charts on the complex projective space other
than~\(z^n=1\), which would lead to different maps with
different convergence properties. A third possibility is to use the
freedom in choosing the diagonal matrix~\(D\)---e.g. by
optimizing the product of norms~\(\Vert \theta\Vert \Vert\Delta\Vert\)---to construct maps
with larger convergence domains, a trick which is known to sometimes
improve the convergence of the RS series~\cite{Remez_2018,Szabados_1999}.

Finally, it would be desirable to obtain sharper theoretical results on
the convergence and speed of our iterative diagonalization algorithm.
This can be difficult: the bifurcations of discrete dynamical systems
are notoriously hard to study in dimensions greater than one, and
estimating convergence rates requires evaluating the Jacobian of the map
at its fixed points---which by definition we do not know~\emph{a
priori}. It is perhaps worth recalling, however, that estimates of radii
of convergence for the RS series were only obtained several decades
after the method was widely applied in continuum and quantum
mechanics~\cite{Kato_1966}, and that their computation in practice
remains an area of active research~\cite{Kvaal_2011}. We leave this
challenge for future work.~ ~

\begin{acknowledgments}
We thank Rostislav Matveev and Alexander Heaton for useful discussions.
Funding for this work was provided by the Alexander von Humboldt
Foundation in the framework of the Sofja Kovalevskaja Award endowed by
the German Federal Ministry of Education and Research.
\end{acknowledgments}

\bibliographystyle{apsrev4-1}
\bibliography{biblio}

\pagebreak

\appendix
\section*{Supplementary Material}
\beginsupplement

\subsection{Eigenvectors and affine
charts}\label{SI_EV}

Let us start with the system of homogeneous polynomial equations $$\mathcal S_n^m:(Mz_n)^mz_n^{\,n}=(Mz_n)^nz_n^{\,m}$$ for all $n$ and $m$. The projective variety defined by this full system is the union of the eigenvectors of $M$ seen as projective points. In other words, the full system $\{\mathcal S_n^m\}_{n,m}$ for all $m$ and $n$ (without assumptions $z^n \neq 0$) is equivalent to the initial eigenvector problem: it is the statement that the exterior product of $Mz$ with $z$ vanishes, which is 
equivalent to $\exists \varepsilon\ Mz = \varepsilon z$.

Assume that $M$ belongs to a perturbative one-parameter family $M = D + \lambda \Delta$. Fix some $n$ and consider the subset $\{\mathcal S_n^m\}_{m}$ of equations that corresponds only to this $n$ and to all $m \neq n$ ($N-1$ equations in total). In the affine chart $U_n \simeq \mathbb C^{N-1}$ defined by $z^n = 1$, this subsystem becomes $z^m = F_n^m(z^k)$ with $F_n^m$ given by \eqref{single_line_dyn}. The fixed points of this dynamical system are in one-to-one correspondence with the eigenvectors of $M$ visible in this chart.

Indeed, we need to prove that for $z \in U_n$ the two following propositions are equivalent: $(i)$ $z$ is a root of $z = F_n(z)$ and $(ii)$ $\exists \varepsilon$ $Mz = \varepsilon z$. The $(ii)$ $\Rightarrow$ $(i)$ direction is proven in the main text. The other direction is easy too. From $z^n = 1$ and $\{\mathcal S_n^m\}_{n,m}$, we conclude that $(Mz)^m = (Mz)^n z^m$ for all $m \neq n$. If we now denote the common factor in these expressions as $\varepsilon = (Mz)^n$, then $(ii)$ immediately follows.

It should be noted, however, that the correspondence holds only in the chart $U_n$. The system $\{\mathcal S_n^m\}_m$ (for the fixed $n$, as before) may have other solutions in $\mathbb CP^{N-1}$, but all of them by necessity are at infinity in $U_n$. These solutions are not related to the eigenvectors of $M$. Indeed, consider $\{\mathcal S_n^m\}_m$ in the form 
\begin{equation}\label{projective}
	z_n^{\,n}z_n^{\,m}=\lambda\theta_{n}^{\,m}\left(z_n^{\,n}(\Delta z_n)^m-z_n^{\,m}(\Delta z_n)^n\right),
\end{equation}
for some fixed $n$ and set $z^n = 0$ to find solutions in $\mathbb CP^{N-1}$ that are at infinity in $U_n$. Clearly, any $z$ with $z^n = 0$ that obeys $(\Delta z)^n = 0$ is a solution of \eqref{projective}. Thus, in general, there is a whole ($N-3$)-dimensional complex projective subspace of such solutions in $\mathbb CP^{N-1}$ (corresponding to a ($N-2$)-dimensional complex hyperplane in the affine space $\mathbb C^N$).

Finally, at $\lambda = 0$ there is only one solution in each chart $U_n$ for each corresponding system $\{\mathcal S_n^m\}_m$, namely the origin of the chart (which represents the corresponding unperturbed eigenvector), while for generic value of $\lambda$, $D$ and $\Delta$ there are all $N$ solutions in each chart for its corresponding system of equations. As soon as $\lambda$ becomes different from $0$, the other $N-1$ solutions appear in the chart from infinity.

\subsection{Rayleigh-Schr\"odinger perturbation
theory}\label{SI_RSPT}

Here we recall the derivation of the Rayleigh-Schr\"odinger recursion \eqref{ARS-alpha}, given \emph{e.g.} in \cite{Roth_2010}. Let $\varepsilon_n$ and $z_n$ respectively be the $n$-th eigenvalue and eigenvector of $M=D+\lambda \Delta$ corresponding to the unperturbed eigenvalues $\epsilon_n$ and eigenvector $e_n$, chosen so that $\lim_{\lambda\to 0}z_n=e_n$ and $\langle e_n,z_n\rangle =1$ for all $\lambda$ (a choice known as ``intermediate normalization''). We start by expanding these in powers of $\lambda$:
$$
\varepsilon_n=\sum_{\ell\geq 0}\lambda^{\ell}\varepsilon_n^{(\ell)}\quad\textrm{and}\quad z_n=\sum_{\ell\geq 0}\lambda^{\ell}z_n^{(\ell)}.
$$
Substituting these ans\"atze into the eigenvalue equation $Mz=\varepsilon z$ and making use of the Cauchy product formula, this yields $Dz_n^{(0)}=\varepsilon_n^{(0)}z_n^{(0)}$ at zeroth order (hence $\varepsilon_n^{(0)}=\epsilon_n$) and for $\ell\geq 1$
$$(D-\epsilon_n)z_n^{(\ell)}=\sum_{s=1}^{\ell}\varepsilon_n^{(s)}z_n^{(\ell-s)}-\Delta z_n^{(\ell-1)}.$$

It is convenient to expand $z_n^{(\ell)}$ in the basis of the eigenstates of $D$ as
$z_n^{(\ell)}=\sum_{m=1}^{N}(a^{(\ell)})_n^{\,m}e_m$ with $(a^{(0)})_n^{\,m}=\delta_{n}^{\, m}$ and $(a^{(\ell)})_n^{\,n}=0$ for $\ell\geq 1$. This gives for each $\ell\geq 1$ and $1\leq m\leq N$
$$
(\epsilon_m-\epsilon_n)(a^{(\ell)})_n^{\,m}=\sum_{s=1}^{\ell}\varepsilon_n^{(s)}(a^{(\ell-s)})_n^{\,m}-(a^{(\ell-1)}\Delta')_{n}^{\,m}.
$$
The equation for the eigenvalues correction is extracted by setting $m=n$ in this equation and using $(a^{(\ell)})_n^{\,n}=\delta_{\ell,0}$. This leads to
$\varepsilon_n^{(\ell)}=(a^{(\ell-1)}\Delta')_{n}^{\,n}.$
Injecting this back into the equation above we arrive at
$$(a^{(\ell)})_n^{\, m}=\theta_{m}^{\, n}\Big(\sum_{s=1}^{\ell} (a^{(s-1)}\Delta')_n^{\, n}(a^{(\ell - s)})_n^{\,m}- (a^{(\ell-1)}\Delta')_n^{\, m}\Big).$$
This is \eqref{ARS-alpha} in the main text.

\subsection{Dynamical perturbation theory contains the RS
series}\label{SI_D_RS_PT}

We prove $A_\mathrm{D}^{(k)}  = A_\mathrm{RS}^{(k)} + \mathcal O(\lambda^{k+1})$ by induction. Obviously $A_\mathrm{D}^{(0)} = A_\mathrm{RS}^{(0)} = I$. Suppose that $A_\mathrm{D}^{(k-1)} = A_\mathrm{RS}^{(k-1)} + \mathcal O(\lambda^k)$ or, more specifically,
$$
A_\mathrm{D}^{(k-1)} = \sum_{\ell = 0}^{k-1} \lambda^\ell  a^{(\ell)} + \mathcal O(\lambda^k),
$$
where the matrices $a^{(\ell)}$ are from the expansion \eqref{ARS-alpha}. Then from \eqref{dyn} we have
$$
A_\mathrm{D}^{(k)} = I +\lambda \theta\star\left( \sum_{\ell=0}^{k-1}\lambda^{\ell} a^{(\ell)}\Delta'-
\left(\sum_{\ell=0}^{k-1} \lambda^\ell a^{(\ell)} \Delta' \right) \triangleright \sum_{m=0}^{k-1} \lambda^m a^{(m)} \right) + \mathcal O(\lambda^{k+1}).
$$
From this expression it is easy to see that the term of $s$-th order in $\lambda$  in $A_\mathrm{D}^{(k)}$ is given by terms with $\ell + m = s - 1$, \emph{viz.}
$$
\lambda^s \left( \theta \star (a^{(s-1)} \Delta') - \theta \star
\left(\sum_{\ell=0}^{s-1} \left(a^{(\ell)} \Delta' \right) \triangleright a^{(s-1-\ell)}\right)\right).
$$
This term matches exactly the RS coefficient $\lambda^s a^{(s)}$ as given by \eqref{ARS-alpha}. This concludes the proof. 

\subsection{Convergence domain of the dynamical perturbation
theory}\label{SI_domain}

Consider a map $F_n$ for some fixed $n$. An attracting equilibrium point of the corresponding dynamical system $z^{(k)}=F_n(z^{(k-1)})$ loses its stability when the Jacobian matrix $(\partial F_n)^m_k \equiv \partial F_n^m/\partial z^k$ of $F_n$ has an eigenvalue (or \emph{multiplier}) with absolute value equal to $1$ at this point.

Consider the system of $N+1$ polynomial equations of $N+2$ complex variables ($z^m$ for $1\leq m\leq N$, $\lambda$, and $\mu$)
$$\begin{cases}
z = F_n(z),\\
\det(\partial F_n - \mu I) = 0.
\end{cases}$$
The variable $\mu$ here plays the role of a multiplier of a steady state. Either by successively computing resultants or by constructing a Groebner basis with the correct lexicographical order, one can exclude the variables $z^m$ from this system, which results in a single polynomial of two variables $(\lambda,\mu) \mapsto P(\lambda,\mu)$. This polynomial defines a complex 1-dimensional variety. The projection to the $\lambda$-plane of the real 1-dimensional variety defined by $\{P=0, \vert\mu\vert^2= 1\}$ corresponds to some curve $C$. A more informative way is to represent this curve as a complex function of a real variable $t$ implicitly defined by $P(\lambda,e^{it}) = 0$.

The curve $C$ is the locus where a fixed point of $F_n$ have a multiplier on the unit circle. In particular, the fixed point that at $\lambda = 0$ corresponds to $z^n = 1$ and $z^m = 0$, $m \neq n$, loses its stability along a particular subset of this curve. The convergence domain of the dynamical perturbation theory is the domain that is bounded by these parts of the curve and that contains $0$.

$C$ is a smooth curve with cusps (return points), which correspond to the values of $\lambda$ such that $M$ has a nontrivial Jordan form (is non-diagonalizable). In a typical case, all cusps a related to the merging of a pair of eigenvectors of $M$. For the dynamical system \eqref{single_line_dyn}, the cusps, thus, correspond to the fold bifurcations of its steady states \cite{DOLOTIN_2008}. One of the multipliers equals to $1$ at such merged point \cite{Kuznetsov_2004}, so these values of $\lambda$ can be found as a subset of the roots of the univariate polynomial $P(\lambda,1)$. Not all its roots generally correspond to cusps and fold bifurcations, though, as we will see later. On the other hand, all the cusps/fold bifurcation points are among the $\lambda$-roots of the discriminant polynomial $\mathrm{Disc}_x\det(M - xI)$. These roots may correspond to both geometric (when two eigenvectors merge and $M$ acquires nontrivial Jordan form) and algebraic (when the eigenvalues become equal but $M$ retains trivial Jordan form) degeneration of $M$. Algebraic degenerations are not related to cusps. Thus, the cusp points of $C$ can be found as the intersection of the root sets of the two polynomials.

The convergence domain of the dynamical system \eqref{dyn} for the whole matrix $A$ of the eigenvectors is equal to the intersection of the convergence domains for its individual lines \eqref{single_line_dyn}.

The explicit $2\times 2$ example of the main text results in the curve given by $4\lambda^2 + e^{it}(2 - e^{it}) = 0$ for each line of the matrix $A$. The cusps of this curve are at $\lambda = \pm i/2$. In this particular case, the convergence circle of the RS perturbation theory is completely contained in the convergence domain of the dynamical perturbation theory, and their boundaries intersect only at the cusp points.

This convergence domain is directly related to the main cardioid of the classical Mandelbrot set: the set of complex values of the parameter $c$ that lead to a bound trajectory of the classical quadratic (holomorphic) dynamical system $x^{(k+1)} = (x^{(k)})^2 + c$. The main cardioid of the Mandelbort set (the domain of stability of a steady state) is bounded by the curve $4c - e^{it}(2 - e^{it}) = 0$. The boundary of the stability domain of our $2\times2$ example is simply a conformal transform of this cardioid by two complementary branches of the square root function composed with the sign inversion. The origin of this relation becomes obvious after the parameter change $c \mapsto -\lambda^2$ followed by the variable change $x \mapsto \lambda x$. This brings the classical system to the dynamical system of the only nontrivial component of the first line for our $2\times 2$ example: $x^{(k+1)} = \lambda \left((x^{(k)})^2 - 1\right)$. The nontrivial component of the second line follows an equivalent (up to the sign change of the variable) equation: $x^{(k+1)} = \lambda \left(1 - (x^{(k)})^2 \right)$.

\subsection{Explicit examples}\label{SI_examples}

The $2 \times 2$ example in the main text is very special. In fact, any 2-dimensional case is special in the following sense. The iterative approximating sequence for $M = D + \lambda \Delta$ for any $D$ and $\Delta$ takes the form
$$A^{(k)}_\mathrm{D} = \begin{pmatrix}
    1 & f_1^{\circ k}(0) \\
    f_2^{\circ k}(0) & 1
\end{pmatrix},$$
where $f_j$ are univariate quadratic polynomials related by $g_1(x) = x^2 g_2(1/x)$, with $g_j(x) \equiv x - f_j(x)$. The first special feature of this recursion scheme is that it is equivalent to a 1-dimensional quadratic discrete-time dynamical system for each line of $A$. This implies that the only critical point of either $f_j$ ($0$, when the diagonal elements of $\Delta$ are equal) is necessarily attracted by at most unique stable fixed point. The second special feature is fact that both lines have exactly the same convergence domains in the $\lambda$-plane. To see this, suppose that $x_1$ and $x_2$ are the roots of $g_1$. Then it follows that $1/x_1$ and $1/x_2$ are the roots of $g_2$. As $g_2'(1/x) = 2g_1(x)/x - g_1'(x)$, and (like for any quadratic polynomial) $g_1'(x_1) = -g_1'(x_2)$, we also see that $g_1'(x_{1,2}) = g_2'(1/x_{2,1})$, and thus $f_1'(x_{1,2}) = f_2'(1/x_{2,1})$. Therefore, the fixed point of the dynamical systems defined by $x^{(k+1)} = f_1(x^{(k)})$ corresponding to an eigenvector and the fixed point of $x^{(k+1)} = f_2(x^{(k)})$ corresponding to the different eigenvector are stable or unstable simultaneously.

\begin{figure}[h!]
\begin{center}
\includegraphics[width=0.70\columnwidth]{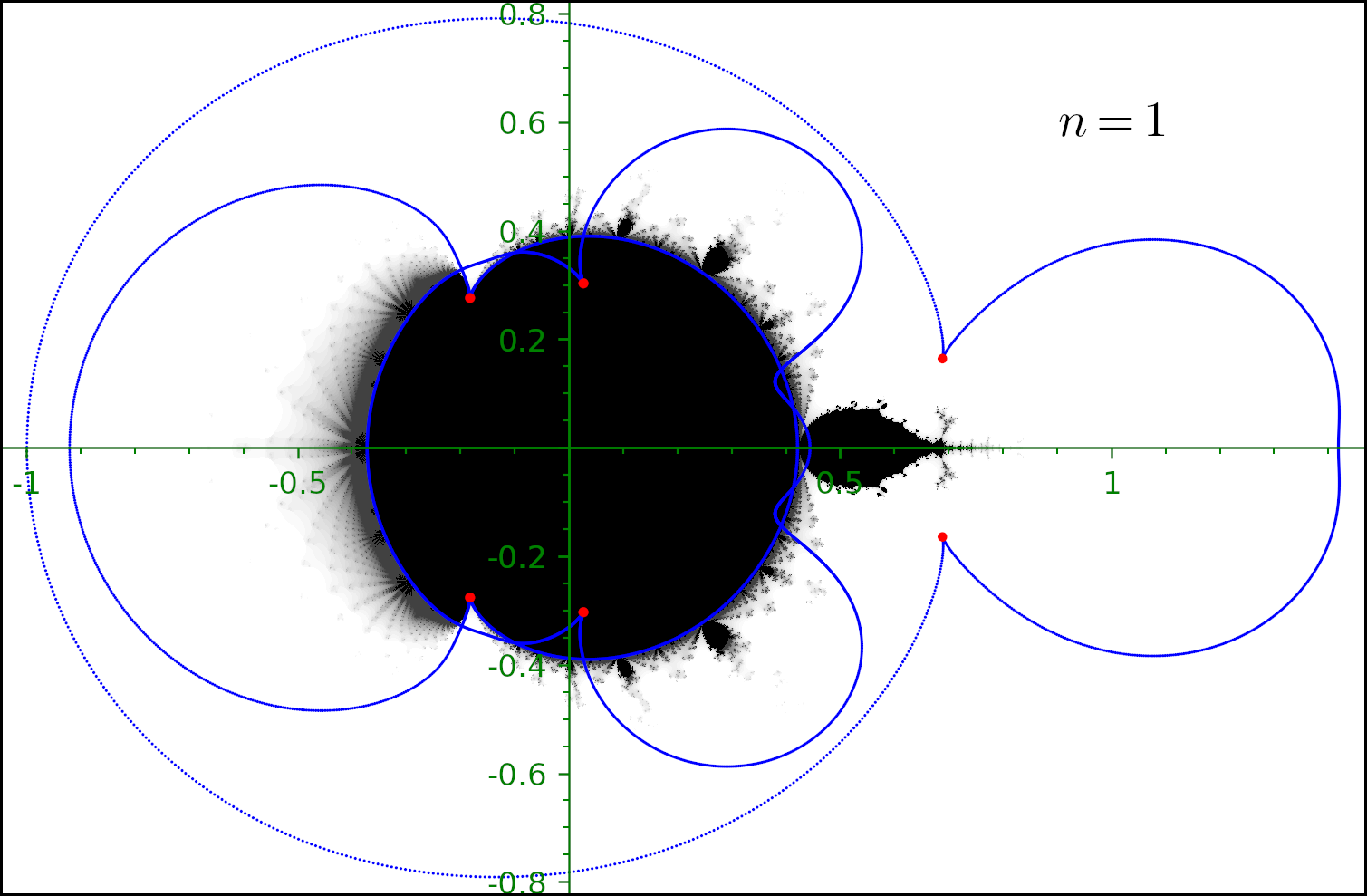}
\caption{The convergence domain on the~\(\lambda\)-plane for the first
line of~\(A\)~(the first eigenvector \(z_1\))
for the~\(3\times3\) example. The Mandelbrot-like set (domain where orbits remain bounded) of the iterative scheme is shown in black and grey. The empirical convergence domain is shown in black. Its largest component corresponds to the stability of a steady state (the applicability domain of the iterative method). Small components correspond to stability of various periodic orbits. Various shades of grey show the values of $\lambda$ that lead to divergence to infinity (the darker the slower the divergence). In red are the values of $\lambda$ where the matrix is non-diagonalizable.
{\label{473965}}%
}
\end{center}
\end{figure}

\begin{figure}[h!]
\begin{center}
\includegraphics[width=0.70\columnwidth]{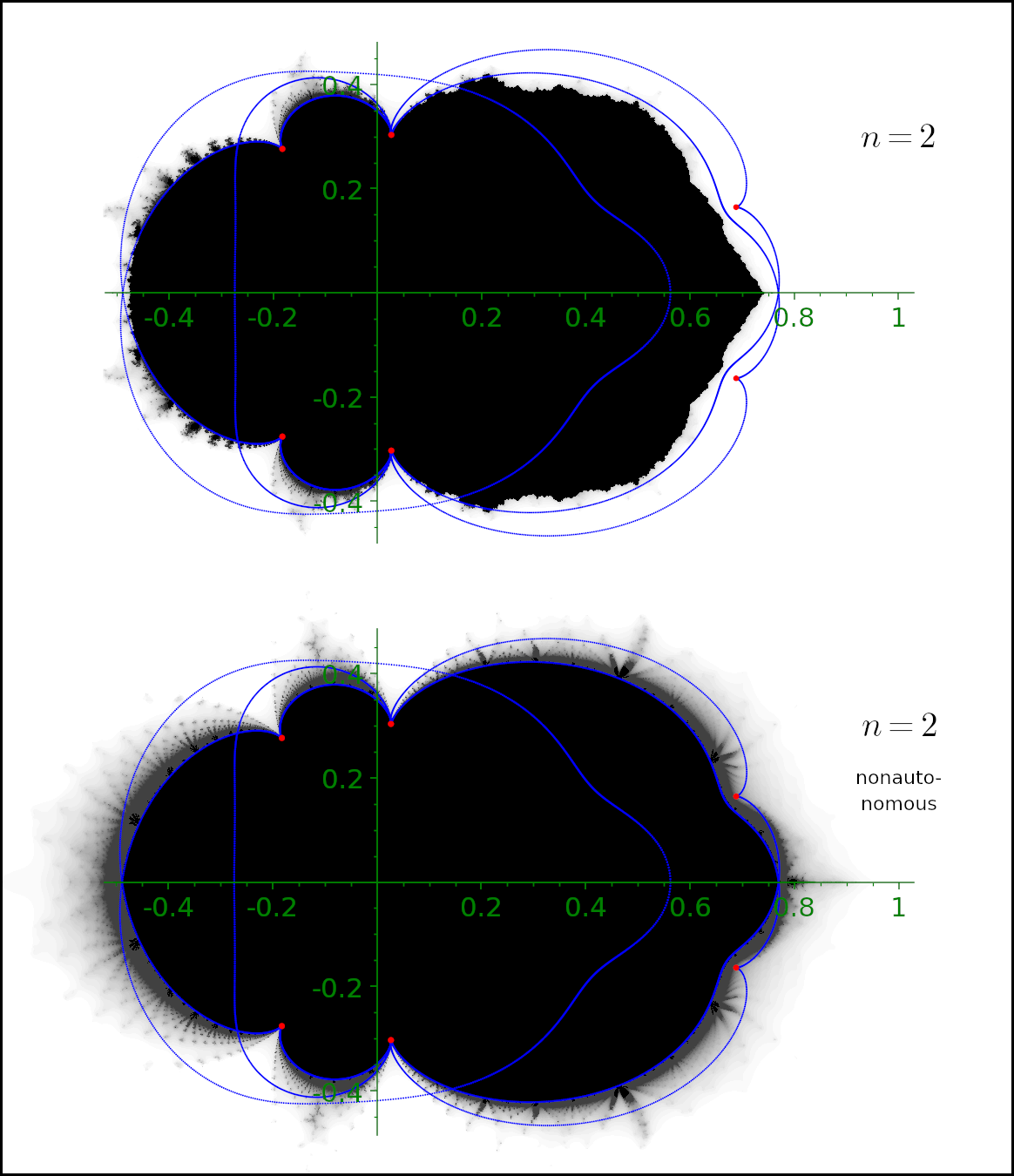}
\caption{{Same as Fig.~{\ref{473965}} for the second
eigenvector~\(z_2\) (the second line of \(A\)).
Small components correspond to stability of various periodic orbits.
Various shades of grey show the values of~\(\lambda\) that lead
to divergence to infinity (the darker the slower the divergence).
{\label{988969}}%
}}
\end{center}
\end{figure}

\begin{figure}[h!]
\begin{center}
\includegraphics[width=0.70\columnwidth]{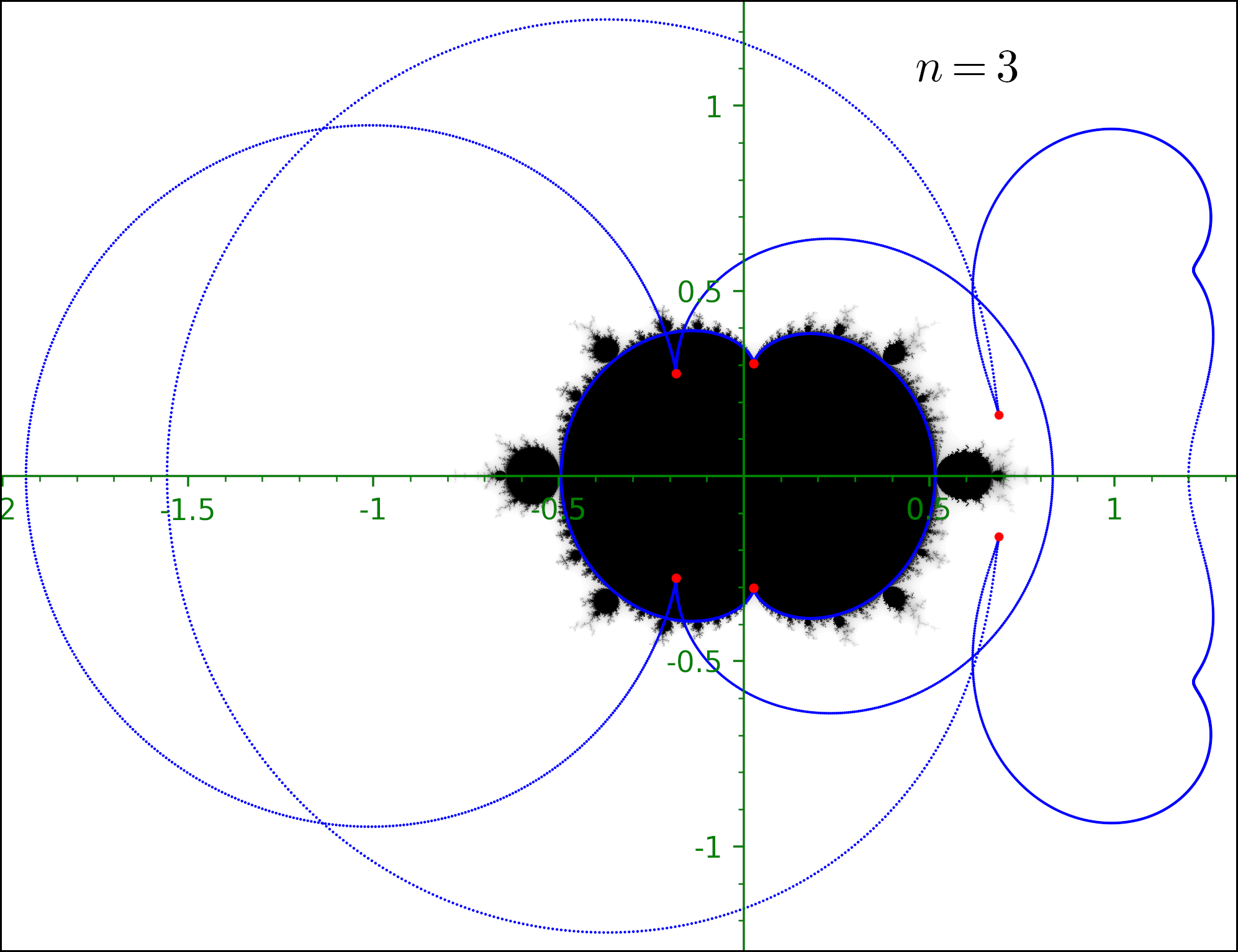}
\caption{{Same as Fig.~{\ref{473965}} for the third
eigenvector~\(z_3\) (the third line of~\(A\)) .
{\label{824432}}%
}}
\end{center}
\end{figure}

These two properties are not generic when $N>2$. Therefore, we provide another explicit example of a $3 \times 3$ matrix to foster some intuition for more general cases:
$$
M = \begin{pmatrix}
0 & 0 & 0 \\
0 & 1 & 0 \\
0 & 0 & 3
\end{pmatrix}
+
\lambda \begin{pmatrix}
0 & 1 & 2 \\
1 & 0 & 3 \\
2 & 3 & 0
\end{pmatrix}.
$$
The polynomial $P$ that defines the fixed point degeneration curve $C$ (see section \ref{981484}) here takes the form for $n = 1$
\begin{align}
P(\lambda,\mu) &= 63792 \lambda^{7} - 28352 \lambda^{6} \mu - 68040 \lambda^{6} - 29556 \lambda^{5} \mu^{2} + 89352 \lambda^{5} \mu\nonumber\\
&- 13239 \lambda^{5} + 960 \lambda^{4} \mu^{3}
 + 14516 \lambda^{4} \mu^{2} - 39164 \lambda^{4} \mu + 12116 \lambda^{4}\nonumber\\ 
&+ 5616 \lambda^{3} \mu^{4} - 26658 \lambda^{3} \mu^{3} + 29988 \lambda^{3} \mu^{2} - 546 \lambda^{3} \mu - 2448 \lambda^{3}\nonumber\\ 
&+ 468 \lambda^{2} \mu^{5} - 3720 \lambda^{2} \mu^{4} + 12820 \lambda^{2} \mu^{3} - 17648 \lambda^{2} \mu^{2} + 7584 \lambda^{2} \mu\nonumber\\ 
&- 1296 \lambda^{2} - 243 \lambda \mu^{6} + 1404 \lambda \mu^{5} - 2619 \lambda \mu^{4} + 1350 \lambda \mu^{3} + 432 \lambda \mu^{2}\nonumber\\ 
&+ 108 \mu^{6} - 792 \mu^{5} + 1980 \mu^{4} - 1872 \mu^{3} + 432 \mu^{2},
\label{poly-n1}
\end{align}
for $n = 2$
\begin{align}
P(\lambda,\mu) &= 113408 \lambda^{7} - 63792 \lambda^{6} \mu - 120960 \lambda^{6} + 7416 \lambda^{5} \mu^{2} + 53208 \lambda^{5} \mu\nonumber\\  
&+ 36424 \lambda^{5} + 6525 \lambda^{4} \mu^{
3} - 11034 \lambda^{4} \mu^{2} - 13824 \lambda^{4} \mu - 5664 \lambda^{4}\nonumber\\ 
&- 3156 \lambda^{3} \mu^{4} - 2472 \lambda^{3} \mu^{3} + 10332 \lambda^{3} \mu^{2} + 3696 \lambda^{3} \mu + 3088 \lambda^{3}\nonumber\\  &- 72 \lambda^{2} \mu^{5} + 1800 \lambda^{2} \mu^{4} - 1800 \lambda^{2} \mu^{3} - 2088 \lambda^{2} \mu^{2} - 1296 \lambda
^{2} \mu\nonumber\\ 
&- 576 \lambda^{2} + 128 \lambda \mu^{6} - 24 \lambda \mu^{5} - 736 \lambda \mu^{4} - 120 \lambda \mu^{3} + 1328 \lambda \mu^{2}\nonumber\\ 
&- 72 \mu^{6} + 108 \mu^{5} + 360 \mu^{4} - 432 \mu^{3} - 288 \mu^{2},
\label{poly-n2}
\end{align}
and for $n = 3$
\begin{align}
P(\lambda,\mu) &= 35440 \lambda^{7} - 42528 \lambda^{6} \mu - 37800 \lambda^{6} - 32360 \lambda^{5} \mu^{2} + 110080 \lambda^{5} \mu\nonumber\\ 
&- 23295 \lambda^{5} + 29112 \lambda^{4} \mu^
{3} - 4800 \lambda^{4} \mu^{2} - 88614 \lambda^{4} \mu + 51024 \lambda^{4}\nonumber\\ 
&+ 14640 \lambda^{3} \mu^{4} - 78760 \lambda^{3} \mu^{3} + 116760 \lambda^{3} \mu^{2} - 52920 \lambda^{3} \mu + 5400 \lambda^{3}\nonumber\\ 
&- 2376 \lambda^{2} \mu^{5} - 10152 \lambda^{2} \mu^{4} + 70296 \lambda^{2} \mu^{3} - 101496 \lambda^{2} \mu^{2} + 41904 \lambda^{2} \mu\nonumber\\ 
&- 864 \lambda^{2} - 1080 \lambda \mu^{6} + 7920 \lambda \mu^{5} - 19620 \lambda \mu^{4} + 19440 \lambda \mu^{3} - 6480 \lambda \mu^{2}\nonumber\nonumber\\ 
&+ 1296 \mu^{6} - 7992 \mu^{5} + 17712 \mu^{4} - 16416 \mu^{3} + 5184 \mu^{2}.
\label{poly-n3}
\end{align}
As we can see, the dynamical systems for all three lines of $A$ ($n = 1, 2, 3$) have different domains of convergence in the $\lambda$ plane. The corresponding curves are depicted on Fig.~\ref{473965}, Fig.~\ref{988969}, and Fig.~\ref{824432}, respectively.

There are differences also in curves for individuals lines with those for the $2\times 2$ case. Note that a curve $C$ does not contain enough information to find the convergence domain itself. The domains on Fig.~\ref{473965}--\ref{824432} were found empirically. Of course, they are bound by some parts of $C$ and include the point $\lambda = 0$. The reason for some parts of $C$ not forming the boundary of the domain is that its different parts correspond to different eigenvectors. In other words, they belong to different branches of a multivalued eigenvector function of $\lambda$, the cusps being the branching points.

Consider as a particular example the case $n = 2$. The curve $C$ intersects itself at $\lambda \approx -0.49$. Above the real axis about this point, one of the two intersecting branches of the curve form the convergence boundary. Below the real axis, the other one takes its place. This indicates that the two branches correspond to different multipliers of the same fixed point. When $\Im \lambda > 0$, one of them crosses the unitary circle at the boundary of the convergence domain;  when $\Im \lambda < 0$, the other one does. At $\lambda \approx -0.49$, both of them cross the unitary circle simultaneously. This situation corresponds, thus, to a Neimark-Sacker bifurcation (the discrete time analog of the Andronov-Hopf bifurcation). Both branches are in fact parts of the same continuous curve that passes through the point $\lambda \approx 0.56$. Around this point, the curve poses no problem to the convergence of the dynamical system. The reason for this is that an excursion around cusps (branching points of eigenvectors) permutes some eigenvectors. As a consequence, the curve at $\lambda \approx -0.49$ corresponds to the loss of stability of the eigenvector that is a continuation of the unique stable eigenvector at $\lambda = 0$ by the path $[0,-0.49]$. At the same time, the same curve at $\lambda \approx 0.56$ indicates a unitary by absolute value multiplier of an eigenvector that is not a continuation of the initial one by the path $[0,0.56]$.

Not all features of the curves depicted this $3\times3$ case are generic either. The particular symmetry with respect to the complex conjugation of the curve and of its cusps is not generic for general complex matrices $D$ and $\Delta$, but it is a generic feature of matrices with real components. In this particular case, due to this symmetry, the only possible bifurcations for real values of $\lambda$ are the flip bifurcations (a multiplier equals to $-1$, typically followed by the cycle doubling cascade), the Neimark-Sacker bifurcation (two multipliers assume complex conjugate values $e^{\pm it}$ for some $t$), and, if the matrices are not symmetric, the fold bifurcation (a multiplier is equal to $1$). With symmetric real matrices, the fold bifurcation is not encountered because the cusps cannot be real but instead form complex conjugated pairs. These features are the consequence of the behavior of $\det(D + \lambda \Delta - x I)$ with respect to complex conjugation and from the fact that symmetric real matrices cannot have nontrivial Jordan forms.

Likewise, Hermitian matrices result in complex conjugate nonreal cusp pairs, but the curve itself is not necessarily symmetric. As a result, there are many more ways for a steady state to lose its stability, from which the fold bifurcation is, however, excluded. Generic bifurcations at $\lambda \in \mathbb R$ here consist in a multiplier getting a value $e^{it}$ for some $t \neq 0$. The situation for general complex matrices lacks any symmetry at all. Here steady states lose their stability by a multiplier crossing the unit circle with any value of $t$, and thus the fold bifurcation, although possible, is not generic. It is generic for one-parameter (in addition to $\lambda$) families of matrices.

As already noted, for the holomorphic dynamics of  any $2\times2$ case the unique critical point is guaranteed to be attracted by the unique attracting periodic orbit, if the latter exists. This, in turn, guarantees that for any $\Delta$ with zero diagonal the iteration of \eqref{dyn} starting from the identity matrix converges to the needed solution provided that $\lambda$ is in the convergence domain. This is not true anymore for $N > 2$, starting already from the fact that there are no discrete critical points in larger dimensions. The problem of finding a good initial condition becomes non-trivial. As can be seen on Fig.~\ref{988969}, the particular $3\times3$ case encounters this problem for the second line ($n = 2$). The naive iteration with $A^{(0)} = I$ does not converge to the existing attracting fixed point of the dynamical system near some boundaries of its convergence domain. Our current understanding of this phenomenon is the crossing of the initial point by the attraction basin boundary (in the $z$-space). This boundary is generally fractal. Perhaps this explains the eroded appearance of the empirical convergence domain of the autonomous iteration.

To somewhat mitigate this complication, we applied a nonautonomous iteration scheme in the form, omitting details, $z^{(k+1)} = F_2(z^{(k)},\lambda (1 - \alpha^k))$ with $z^{(0)} = (0,1,0)$, where $\alpha$ is some positive number $\alpha < 1$, so that $\lim_{k \to \infty} \lambda (1 - \alpha^k) = \lambda$, and we explicitly indicated the dependence of $F_2(z,\lambda)$ on $\lambda$. The idea of this \emph{ad hoc} approach is the continuation of the steady state in the extended $(z,\lambda)$-phase space from values of $\lambda$ that put $z^{(0)}$ inside the convergence domain of that steady state. Doing so, we managed to empirically recover the theoretical convergence domain (see Fig.~\ref{988969}).

Finally, we would like to point out an interesting generic occurrence of a unitary multiplier without the fold bifurcation. For $n = 1$, this situation takes place at $\lambda \approx 0.45$, for $n = 2$ at $\lambda \approx 0.56$, and for $n = 3$ at $\lambda \approx 1.2$. All three points are on the real axis, as  is expected from the symmetry considerations above. There is no cusp at these points and no fold bifurcations (no merging of eigenvectors), as it should be for symmetric real matrices. Instead, another multiplier of the same fixed point goes to infinity at the same value of $\lambda$ (the point becomes super-repelling). As a result, the theorem of the reduction to the central manifold is not applicable.

\end{document}